\documentclass[notitlepage,showpacs,aps,prl,onecolumn,showkeys,nofootinbib,nobibnotes]{revtex4-1}
\usepackage[dvipdfmx]{graphicx}
\usepackage{url}

\usepackage{lineno}
\modulolinenumbers[5]
\newcounter{bla}

\usepackage{graphicx,color}
\usepackage{amssymb,amsfonts,amsmath}
\usepackage[dvipdfmx]{hyperref}
\title{}
\begin{document}

\title{Deterministic replica-exchange method without pseudo random numbers for simulations of complex systems }


\author{ Ryo Urano$^{\rm 1}$ and Yuko Okamoto$^{{\rm 1,2,3,4}}$ } 
\affiliation{$^1$Department of Physics, Graduate School of Science, Nagoya University, Nagoya, Aichi 464-8602, Japan}    
 \affiliation{$^{\rm 2}$Structural Biology Research Center, Graduate School of Science, Nagoya University, Nagoya, Aichi 464-8602, Japan} 
\affiliation{$^{\rm 3}$Center for Computational Science, Graduate School of Engineering, Nagoya University, Nagoya, Aichi 464-8603, Japan}
\affiliation{$^{\rm 4}$Information Technology Center, Nagoya University, Nagoya, Aichi 464-8601, Japan
}





\begin{abstract}
We propose a replica-exchange method (REM) which does not use pseudo random numbers. For this purpose, 
we first give a conditional probability for Gibbs sampling replica-exchange method (GSREM) based on the heat bath method.
In GSREM, replica exchange is performed by conditional probability based on the weight of states using pseudo random numbers.
From the conditional probability, we propose a new method called deterministic replica-exchange method (DETREM) that produces thermal equilibrium distribution based on a differential equation  instead of using pseudo random numbers.  
This method satisfies the detailed balance condition using a conditional probability of Gibbs heat bath method and thus results can reproduce the Boltzmann distribution within the condition of the probability.
We confirmed that the equivalent results were obtained by REM and  DETREM with two-dimensional Ising model. DETREM can avoid problems of choice of seeds in pseudo random numbers for parallel computing of REM and gives analytic method for REM using a differential equation.
\end{abstract}


\maketitle


\section*{Introduction}
\label{sec-1}
            The enhancement of sampling during Monte Carlo (MC) and molecular dynamics (MD) simulations is very important for complex systems. Replica-exchange method (REM) (or parallel tempering) is one of the most popular ways to improve sampling efficiency\cite{rem1,swendsen1986replica,rem3,rem4} including biomolecular system in explicit solvent\cite{liu2005replica,wang2011replica}  or biomembrane\cite{mori2013surface,mori2014surface} (for reviews, see, e.g., Refs.\cite{mitsutake_generalizedensemble_2001,rem_iba2001extended}).
To realize a thermal equilibrium distribution, REM uses Metropolis criterion with pseudo random numbers. However, 
random numbers sometimes give inaccurate results of simulations\cite{georgescu2012locally}. 
Moreover, generation of high quality random numbers is often difficult and does not assure good simulation results\cite{ferrenberg1992monte}. 
REM and its extension is suited for parallel computing\cite{sugita2012recent,vogel2013generic,vogel2014scalable,vogel2014exploring}. Most of pseudo random number generators decrease the scalability in parallelization\cite{salmon2011parallel}.
Hence, the complementary method producing the same results without pseudo random numbers is meaningful. 

In addition, the analytic approach for temperature selections in REM have been performed \cite{trebst2004optimizing,nadler2007dynamics}.
For performance and the condition of REM, several works were also performed.
For example, Nymeyer \cite{nymeyer2008efficient} showed how efficient REM is than conventional simulations using the number of independent configurations.
Abraham and Gready introduced some measurement and compared the results\cite{abraham2008ensuring}.
Rosta and Hummer \cite{rosta2009error} evaluated the practical efficiency of REM simulation for protein folding with a two-state model.
However, the examination of the condition for convergence of REM is difficult partly because the mixing of temperature in REM is determined by pseudo random numbers with Metropolis criteria. As a result, most of analyses estimated the REM performance by simulation results. 

Recently, Suzuki $et\ al$. proposed a method to produce a thermal equilibrium state without using random numbers for spin models by a differential equation based on the conditional probability of Gibbs sampling heat bath method, which is referred to as chaotic Boltzmann machines\cite{suzuki2013chaotic,suzuki2013monte}.
 The differential equation controls spin states at each site and the staying time of each spin state is proportional to the weights of the thermal equilibrium distribution.
 They reproduced the results of a conventional MC method in some spin systems.

Moreover, Boltzmann machine \cite{ackley1985learning} has its mathematical framework \cite{wainwright2008graphical}.
The method was analyzed by mean field approximation\cite{tanaka1998mean}, algebraic geometry and informative geometry. 
For example, a linear convergence of parameters in Boltzmann machine was suggested by a learning algorithm of Fisher information matrices\cite{amari1992information}, and a upper boundary for performance was obtained by algebraic geometry \cite{yamazaki2005singularities}.
By introducing the differential equation for replica-exchange method,  the previous results in the fields can be applied for the REM analysis. 
This means that analytic approach for Boltzmann machine will be applied for REM by this extension.
Moreover, this new implementation of REM will be related to hierarchical structure of Boltzmann machine, which is similar to deep Boltzmann machine \cite{salakhutdinov2009deep,salakhutdinov2012efficient}. Developments in Boltzmann machine to accelerate convergence of sampling  such as Contrastive Divergence method\cite{hinton2002training} have been proposed.

We here generalize this Chaotic Boltzmann machine to REM.
We first have to extend the conditional probability for replica exchange not based on Metropolis criterion but on a Gibbs sampling heat bath method.
The heat bath formalism has already been given in Ref. \cite{chodera2011replica}, we refer to this method as Gibbs sampling replica-exchange method (GSREM). (A similar approach
  based on global balance condition\cite{suwa2010markov} was also developed\cite{global_itoh_replica-permutation_2013}.)
We then introduce a differential equation for replica exchange to modify GSREM. 
This method is referred  to as the deterministic replica-exchange method (DETREM).
We then tested the effectiveness of DETREM by comparing the results of simulation of 2-dimensional Ising model with those by the conventional REM. 

The organization of this paper is as follows.
In Section 2, the theory for the new method and conventional REM is presented.
In Section 3, we give the results of DETREM together with REM.
The final section is devoted to conclusions.

\section*{Methods}
\label{sec-2}

We first briefly review the conventional REM.
We prepare $M$ non-interacting replicas at $M$ different
 temperatures. 
Let the label $i$ (=1, $\cdots$, $M$) stand for the replica index
 and label $m$  (=1, $\cdots$, $M$) for the temperature index. 
Here, $i$ and $m$ are related by the permutation functions by
\begin{eqnarray}
\label{fpermu}
\begin{cases}
i=i(m) \equiv f(m), \\
m=m(i)  \equiv f^{-1} (i),  
\end{cases}
\end{eqnarray}
where $f(m)$ is a permutation function of $m$ and $f^{-1}(i)$ is the inverse.
We represent the state of the entire system of $M$ replicas by $X = \left\{x_{m(1)}^{[1]}  , \cdots, x_{m(M)}^{[M]} \right\}$, where $x_m^{[i]} =\left\{q^{[i]}, p^{[i]}\right\}_m$ are the set of coordinates $q^{[i]}$ and momenta $p^{[i]}$ of particles in replica $i$ (at temperature $T_m$).
The probability weight factor for state $X$ is given by a product of Boltzmann factors: 
\begin{eqnarray}
W_{\rm REM}(X)&=\displaystyle \prod_{i=1}^M \exp{[-\beta_{m(i)} H(q^{[i]} , p^{[i]})]},
\end{eqnarray}
where $\beta_m (=1/k_{\rm B} T_m)$ is the inverse temperature and $H(q,p)$ is the Hamiltonian of the system. 
We consider exchanging a pair of replicas $i$ and $j$
 corresponding to temperatures $T_m$ and $T_n$, respectively: 
\begin{equation}
\label{lstate}
 X =  \left\{ \cdots, x_m^{[i]}  , \cdots,
 x_n^{[j]}, \cdots \right\} \rightarrow   X^\prime =  \left\{ \cdots, x_m^{[j]^\prime}  , \cdots,
 x_n^{[i]^\prime}, \cdots \right\}, 
\end{equation}
where $x_n^{[i]^\prime} \equiv \left\{q^{[i]}, p^{[i]^\prime}\right\}_n ,x_m^{[j]^\prime} \equiv \left\{q^{[j]}, p^{[j]^\prime}\right\}_m$, and $p^{[j]^\prime}=\sqrt{\frac{T_m}{T_n}} p^{[j]},p^{[i]^\prime}=\sqrt{\frac{T_n}{T_m}} p^{[i]}$ \cite{rem4}.
The exchange of replicas introduces  a new permutation function $f^\prime$:
\begin{eqnarray}
\label{newfpermu}
\begin{cases}
i=f(m) \rightarrow j= f^\prime (m), \\
j=f(n)  \rightarrow i= f^\prime (n).  
\end{cases}
\end{eqnarray}
We remark that this process is equivalent to
exchanging a pair of temperatures $T_m$ and $T_n$ for the corresponding replicas $i$ and $j$.

Here, the transition probability $\omega (X\rightarrow X^\prime)$ of Metropolis criterion is given by 
\begin{eqnarray}
\label{metro}
 \omega (X\rightarrow X^\prime )  = {\rm min}\left(1, \frac{W_{\rm REM} (X^\prime)}{W_{\rm REM} (X)}\right)  = {\rm min}(1,\exp(- \Delta  )) ,
\end{eqnarray}
 where 
\begin{eqnarray}
\label{ddelta}
\Delta =\Delta_{m,n}   = (\beta _n - \beta_m ) ( E(q^{[i]}) - E(q^{[j]})   ). 
\end{eqnarray}

REM is performed by repeating the following two steps:

\begin{enumerate}
\item We perform a conventional MD or MC simulation of replica $i\ (=1, \cdots, M)$ at temperature $T_m\ (m=1, \cdots,M)$ simultaneously and independently for short steps.
\item Selected pairs of replicas are exchanged based on the above Metropolis criterion in Eqs. (\ref{metro}) and (\ref{ddelta}). A pseudo random number is used to judge the criterion.
\end{enumerate}
 Without loss of generality we can assume $T_1 < T_2 < \cdots < T_M$. 
Note that in Step 2 we usually exchange only pairs of replicas corresponding to neighboring temperatures, because the acceptance probability for replica exchange decreases exponentially with the difference of the two inverse temperatures and potential energy terms because of Eq. (\ref{ddelta}). 
This replica exchange can be written as
\begin{equation}
\label{lnstate}
 X =  \left\{ \cdots, x_m^{[i]}  , \cdots,
 x_{m+1}^{[j]}, \cdots \right\} \rightarrow   X^\prime =  \left\{ \cdots, x_m^{[j]^\prime}  , \cdots,
 x_{m+1}^{[i]^\prime}, \cdots \right\}, 
\end{equation}
where in Eq. (\ref{metro}) $\Delta$ is now given by
\begin{eqnarray}
\label{nddelta}
\Delta_m   = (\beta _{m+1} - \beta_m ) ( E(q^{[i]}) - E(q^{[j]})   ).
\end{eqnarray}
The REM method makes  a random walk in temperature space during the simulation.
The canonical ensemble is reconstructed by the multiple-histogram reweighting technique, or weighted histogram analysis method (WHAM)\cite{ferrenberg_optimized_1989,kumar_weighted_1992}.

We next present GSREM\cite{chodera2011replica}.
As in the conventional REM, we usually consider the neighboring temperature exchange in Eq. (\ref{lnstate}).
The conditional probability $\omega(x_m^{[j^\prime]} , x_{m+1}^{[i^\prime]} \mid {x^{[k\ne i(m),j(m+1)]}_{m(k)}}) $, in which the new state selects the temperature exchanged state of replicas $i$ and $j$ with $T_{m+1}$ and $T_m$ from the no-exchange state of replicas $i$ and $j$ with temperatures $T_m$ and $T_{m+1}$, is given by
\begin{eqnarray}
\label{gibbsp}
\displaystyle \omega(x_m^{[j^\prime]} , x_{m+1}^{[i^\prime]} \mid {x^{[k\ne i(m),j(m+1)]}_{m(k)}})  &= &\cfrac{ W(x_m^{[j^\prime]} , x_{m+1}^{[i^\prime]} \mid {x^{[k\ne i(m),j(m+1)]}_{m(k)}}) }{
\scriptstyle W(x_m^{[i]} , x_{m+1}^{[j]} \mid {x^{[k\ne i(m),j(m+1)]}_{m(k)}}) +  W(x_m^{[j^\prime]} , x_{m+1}^{[i^\prime]} \mid {x^{[k\ne i(m),j(m+1)]}_{m(k)}})}\\
&=& \cfrac{ 1}{ 1  +\cfrac{W(x_m^{[i]} , x_{m+1}^{[j]} \mid {x^{[k\ne i(m),j(m+1)]}_{m(k)}})}{  W(x_m^{[j^\prime]} , x_{m+1}^{[i^\prime]} \mid {x^{[k\ne i(m),j(m+1)]}_{m(k)}})}} .
\end{eqnarray}

In GSREM, the above procedure for the conventional REM is performed, where Step 2 for the GSREM is performed based on Eq. (\ref{gcond}).
Here, in Step 2,  the conditional probability of a temperature set based on Eq. (\ref{gcond}) is calculated, and this assigns weights between 0 and 1 for exchanged states and a no-exchange state. Finally, after a pseudo random number is generated, the state corresponding to the random number with the assigned region is selected.
For the Boltzmann distribution, this equation in Eq. (\ref{gibbsp}) can be rewritten as 
\begin{equation}
\label{gibbs}
\omega(x_m^{[j^\prime]} , x_{m+1}^{[i^\prime]} \mid {x^{[k\ne i(m),j(m+1)]}_{m(k)}})  = \frac{1}{1+{\rm exp}(\Delta_m)}, 
\end{equation}
where $\Delta_m$ is given by Eq. (\ref{nddelta}).
This is the Gibbs sampling replica-exchange method when an equilibrium state is produced by this conditional probability with pseudo random numbers. 
We remark that REM and GSREM are mathematically equivalent in the present case because updates of new states have only two possibilities.A more general formulation for GSREM is given in the Appendix, which is mathematically different from REM..
We next propose DETREM.
At first, as in the conventional REM, we can only use the
 internal states $y_{m} \in \left\{-1,1\right\}$  for a pair of neighboring temperatures $(T_{m},T_{m+1})$, where the number of internal
 states is $M$-1 with the following pairs: $y_1=(T_1,T_2),y_2=(T_2,T_3), \cdots, y_{M-1}=(T_{M-1},T_M$).
We also propose the differential equation based on Eq. (\ref{gibbs}) given by
\begin{equation}
\label{timeinte}
\frac{d y_{m} }{dt}= \frac{1}{1+ \rm exp(\Delta_m)}, 
\end{equation}
 where $\Delta_m$ is given by Eq. (\ref{nddelta}). 

\begin{figure}[htb]
\centering
\includegraphics[width=1.0\textwidth]{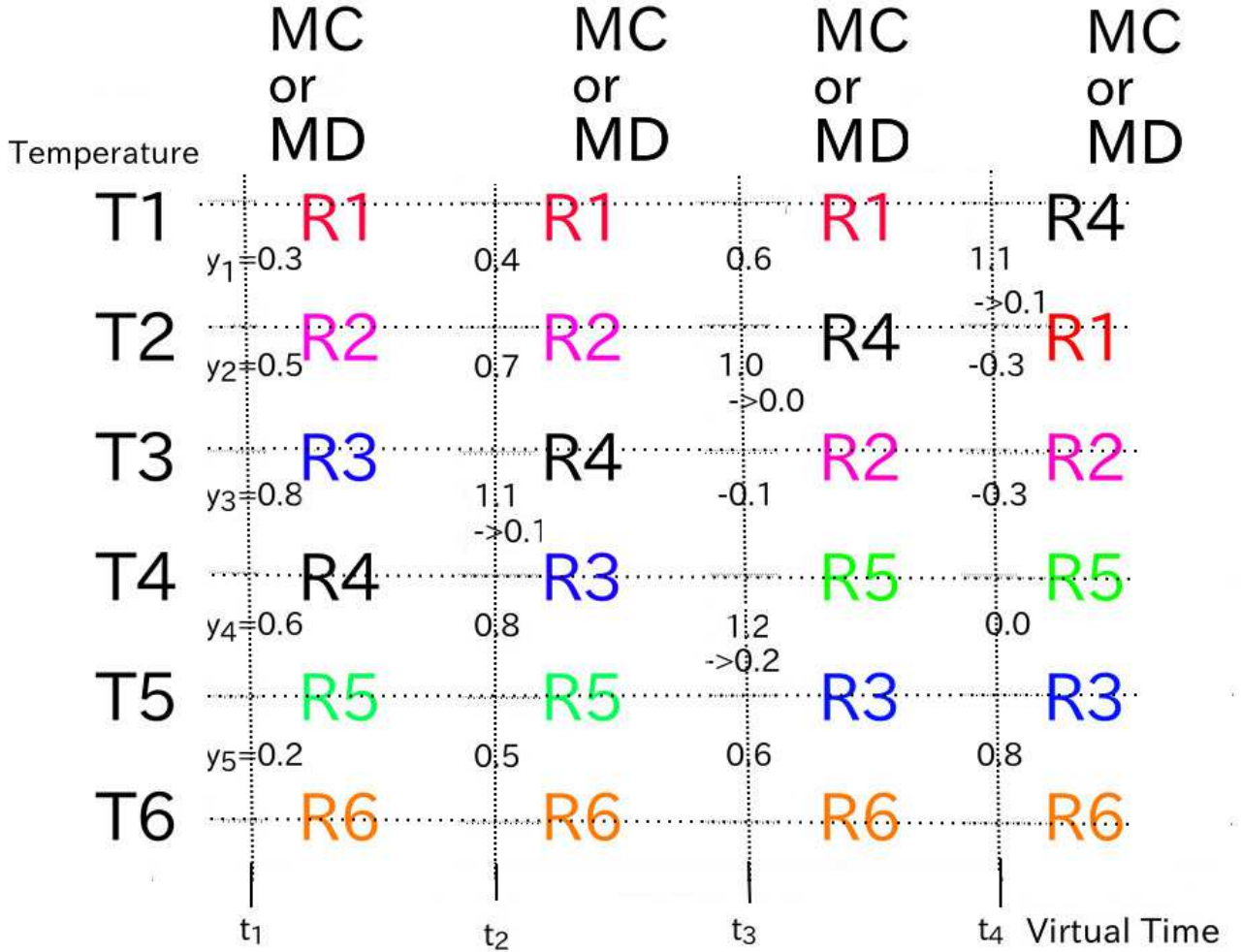}
\caption{\label{scheme}Schematic sketch of DETREM. After parallel conventional MC or MD simulations at $M$ (=6, here) different temperatures for short steps, all internal states y$_m$ are updated. The temperature pairs $(T_m, T_{m+1})$, or corresponding replica pairs, are exchanged when $y_m \ge 1\ {\rm or} \le -1$. The cycle is repeated until the end of simulation. T1, T2, T3, T4, T5, and T6 are temperatures. R1, R2, R3, R4, R5, and R6 are replicas.}
\end{figure}

Compared to REM, the difference of the algorithms is in Step 2. In Step 2 of DETREM, instead of evaluating the Metropolis criterion, the update of each $y_{i}$ is done by the differential equation in Eq. (\ref{timeinte}).
Step 2 in  DETREM is given as follows:
\renewcommand{\labelenumi}{(\roman{enumi})}
\begin{enumerate}
\item All internal states $y_m \ (m=1, \cdots, M-1)$ for replica exchange pairs are integrated independently based on Eq. (\ref{timeinte}).  Namely,  we update $y_m$ by
\begin{equation}
\label{timeconst}
y_m(t+\Delta t) = y_m(t) + \sigma _{m}\frac{1}{1+ \rm exp(\Delta_m)}  dt, \ \ \  \ \ ( m=1, 2, \cdots, M-1)  
\end{equation}
 where $\Delta_m$ is evaluated with the last coordinates in the simulations in Step 1 above and the signature $\sigma_{m} $ of the pair of $(T_m,T_{m+1})$  changes to 1 or $-$1 to control the signature for numerical accuracy of the change of $y_{m}$ which monotonically increases or decrease.
\item When the value of the internal state $y_m$ is $\ge 1$ or $\le -1$, the temperature exchange of the pair corresponding to $y_m$ is performed as follows,  
\begin{eqnarray*}
\displaystyle 
&& {\rm Temperature \ pair\ is\ exchanged}: \\
&& {\rm if\ updated} \ y_m \ge 1, {\rm then}\   {(T_{m}, T_{m+1} ) \rightarrow (T_{m+1}, T_{m} )}, \ y_m \leftarrow y_m -1, \sigma _m \leftarrow -1,   \\
&& {\rm if\ updated} \  y_m \le -1,{\rm  then}\  {(T_{m}, T_{m+1} ) \rightarrow (T_{m+1}, T_{m} )}, \ y_m \leftarrow y_m +1, \sigma _m \leftarrow +1 .
\end{eqnarray*}
 \end{enumerate}
Figure~\ref{scheme} summarizes the algorithm. 
Note that we add value 1 or --1 for new $y_m$ after exchanges are made because of our coding. 
We remark that DETREM is performed just like GSREM, where in Step 2 the evaluation of the conditional probability in Eq. (\ref{gibbs}) by pseudo random numbers is replaced by solving the differential equation in Eq. (\ref{timeinte}).

The DETREM equation is proposed to yield the distributions 
in Eq.~(\ref{gibbs}) based on the conditional probability 
in the heat-bath method. 
While spins are updated in the Chaotic Boltzmann 
machine\cite{suzuki2013monte}, temperatures are updated
in DETREM. The neighboring spins of the chosen spin site
are assumed to be
fixed in the former method, and this assumption for
DETREM corresponds to the one that temperatures in 
unselected replicas are fixed during Step 2 above,
although spins are changed during Step 1.
We remark that DETREM worked properly even when
we changed many spins in Step 1, while only one spin was
updated in the present simulation\cite{urano2015designed}.

Expectation values of physical quantities are given as functions of temperatures by WHAM\cite{ferrenberg_optimized_1989,kumar_weighted_1992,wham3}. 
Namely, the density of states $n(E)$ and dimensionless Helmholtz free energy are obtained by solving the following equations self-consistently:
\begin{equation}
n(E) = { \frac{\sum\limits_{m=1}^{M}N_m(E)}{\sum\limits_{m=1}^{M}n_m e^{f_m-\beta _m E}}},
\end{equation}
and
\begin{equation}
 e^{-f_m}= \sum_{E}n(E) e^{-\beta _m E},
\end{equation}
where $N_m (E)$ and $n_m$ are the energy histogram and the total number
of samples obtained at temperature $T_m$, respectively.
After we obtained $f_m$ at each temperature, 
the expectation value of a physical quantity $A$ at any temperature $T$ is given by \cite{wham3}
\begin{equation}
 <A>_T = \frac{\sum\limits_{m=1}^{M}   \sum\limits_{x_m} A(x_m)
 \displaystyle \frac{1}{\sum\limits_{l=1}^{M}  n_l \exp{(f_l - \beta _l
  E(x_m))}  }  \exp{(-\beta E(x_m))}  }
{ \sum\limits_{m=1}^{M}   \sum\limits_{x_m} 
  \displaystyle \frac{1}{\sum\limits_{l=1}^{M}  n_l \exp{(f_l - \beta _l
  E(x_m))}  }  \exp{(-\beta E(x_m))}  },
\label{wham}
\end{equation}
where $x_m$ are the set of coordinates at temperature $T_m$ obtained from the trajectories of the simulation.

\subsection*{Simulation conditions}
\label{sec-2-1}
In order to test the effectiveness of the present methods, we studied the 2-dimensional Ising  model. The lattice size $L$ in square lattice was 128. The system size $N$ is equal to $L^2$. 
In both methods, the update of spin states was performed by the Metropolis criterion.

For REM, replica-exchange attempt was made for every 1 MC step.   1 MC step was defined to be one Metropolis update of spins. The total number of MC steps was 100,000,000. Each data was sampled at 1,000 MC steps frequency.
To integrate Eq. (\ref{timeinte}), we used the fourth-order Runge-Kutta method, which is equivalent to Eq. (\ref{timeconst}), with virtual time step $dt=1$. 
The initial value for $y_m(t=0)$ was set to 0 for all $m$, while any initial value between $-1$ and 1 is acceptable. Because the initial values of $y_m$ influence only the first replica exchange, the results will not depend on them as long as the total number of replica exchange is large enough. 
The total number of replicas was 40 and the temperatures were 1.50, 1.55, 1.60, 1.65, 1.70, 1.75, 1.80, 1.85, 1.90, 1.94, 1.98,  2.01, 2.04, 2.07, 2.10, 2.13, 2.16, 2.19, 2.22, 2.25, 2.28, 2.31, 2.34,  2.358, 2.368, 2.38, 2.40, 2.42, 2.44, 2.47, 2.51,  2.57, 2.63, 2.69, 2.75, 2.82, 2.90,  3.00, 3.10, and 3.15. Boltzmann constant $k_{\rm B}$ and the coupling constant $J$ were set to 1. Thus, $\beta = 1/ k_{\rm B} T = 1/T =\beta ^*$, and the (potential) energy is given by
 \begin{equation}
E({\bf s})= - \sum_{<i,j>} s_i s_j, 
 \end{equation}
where $s_i=\pm 1$, and the summation is taken over all the nearest-neighbor pairs in the square lattice.
The canonical distribution is given by 
\begin{equation}
 W({\bf s)} =  \frac{1}{Z} {\rm exp}(- \beta^* E({\bf s} ) ), 
\end{equation}
where $Z$ is the partition function. 
In the DETREM simulation,  all $y_m$ were updated simultaneously. 
The multiple exchanges of temperatures at a replica were prohibited.
Namely, if the neighboring internal states $y_m$ and $y_{m+1}$ satisfy the exchange condition ( $y_m \geq 1$ or $y_m \leq$  $-1$), only one state, e.g., $y_m$, was updated and only the pair ($T_m$, $T_{m+1}$)  was exchanged, while $y_{m+1}$ was not updated.
For analysis, we used the R program package.\cite{Rpackage2013,ihak:gent:1996,venables2002modern}

\section*{Results}
\label{sec-3}

\begin{figure}[htb]
\centering
\includegraphics[width=1.0\textwidth]{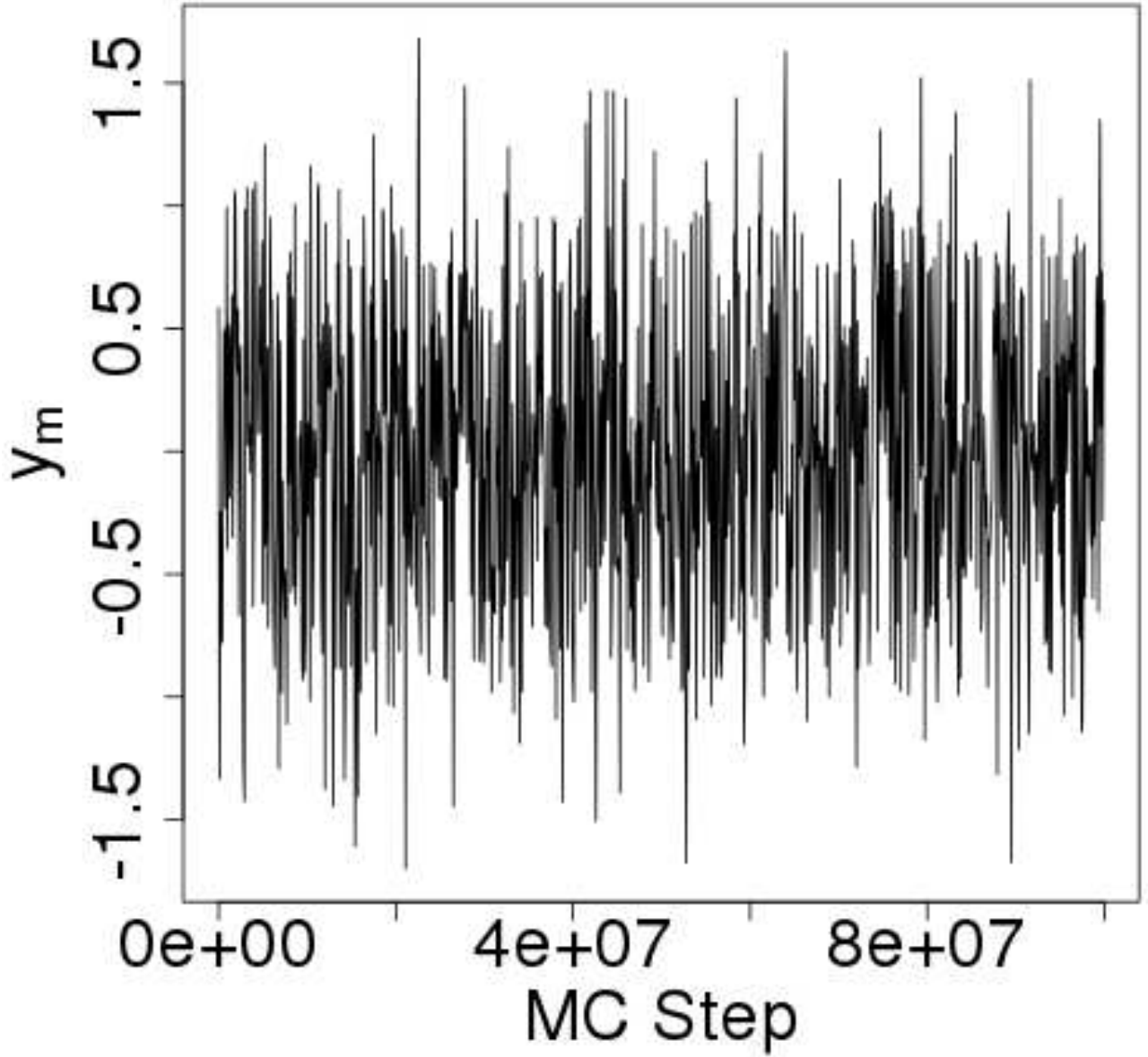}
\caption{\label{ym}Time series of $y_m$ change in one of  $m$ ( $m=1$ ) for DETREM simulations.Replica exchange is performed when $y_m > 1.0$ and $y_m <-1.0$, and $y_m$ is reset to $y_m=y_m -1.0$ and  $y_m=y_m +1.0$, respectively,}
\end{figure}

\begin{figure}[htb]
\centering
\includegraphics[width=1.0\textwidth]{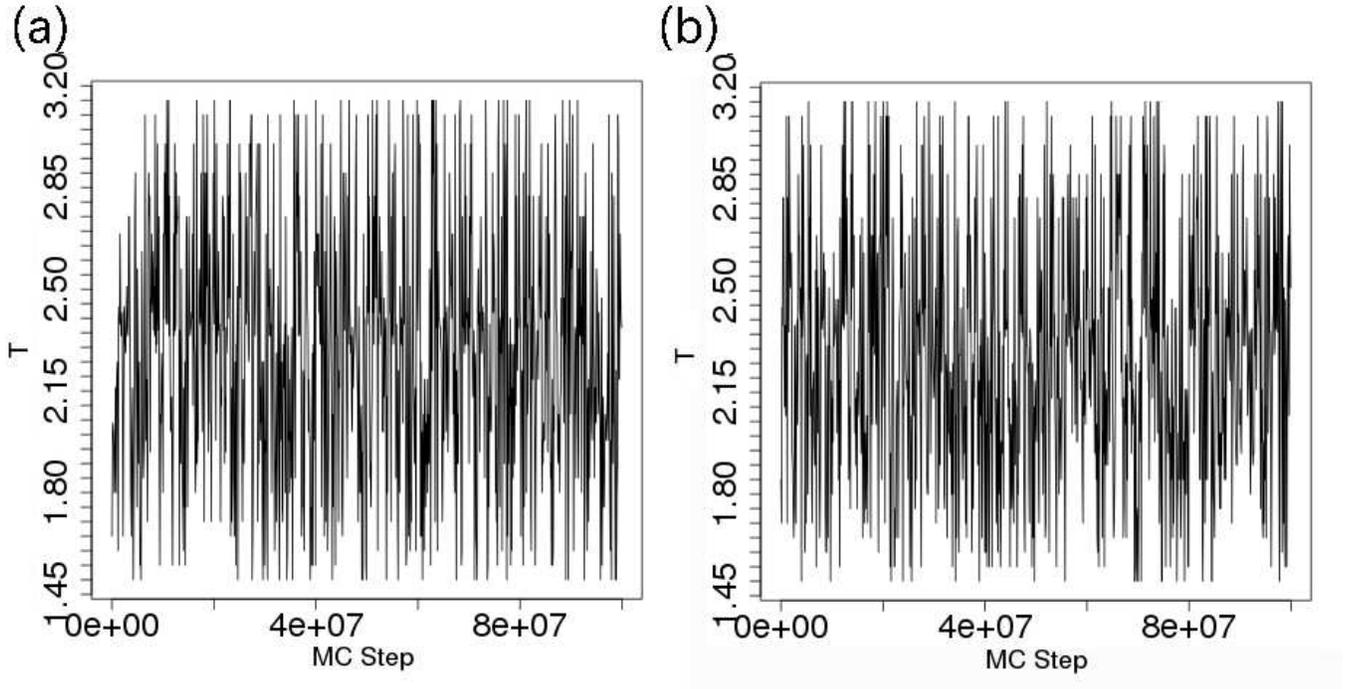}
\caption{\label{rw}Time series of temperature change in one of  the replicas (Replica 1) for (a) REM and (b) DETREM simulations.}
\end{figure}

\begin{figure}[htb]
\centering
\includegraphics[width=1.0\textwidth]{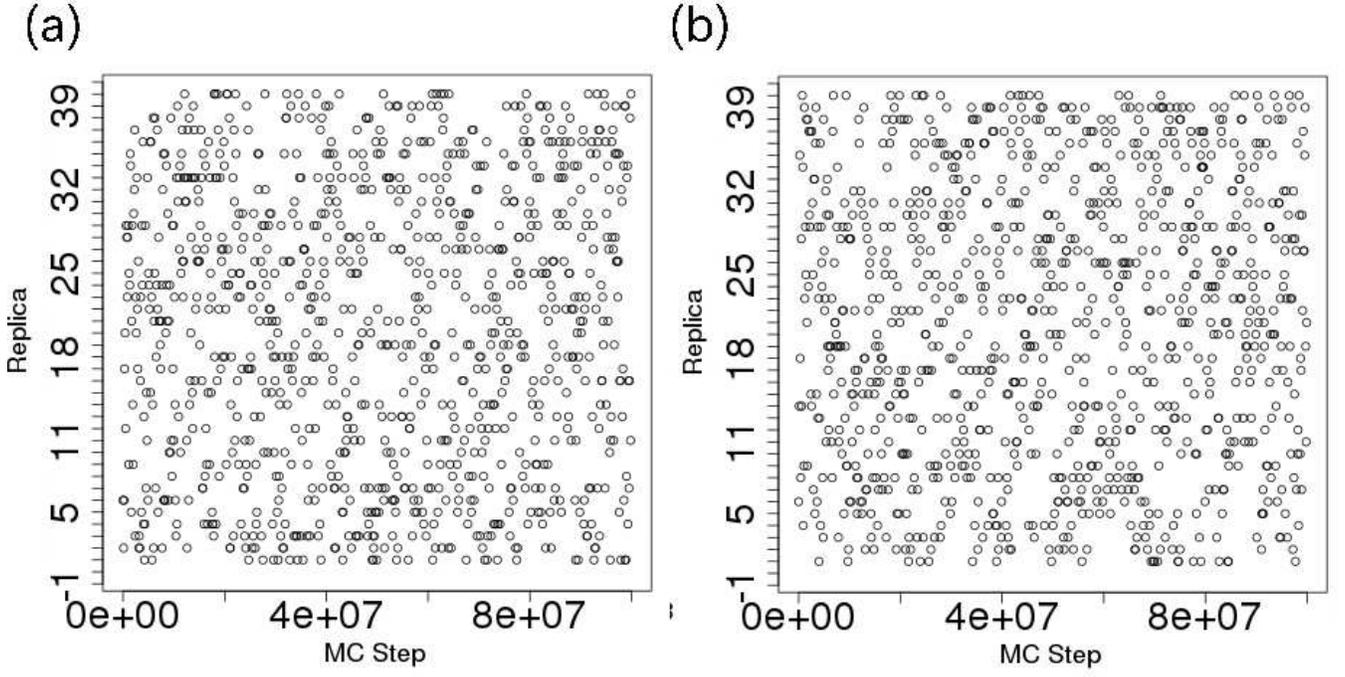}
\caption{\label{ind}Time series of replica at temperature 1.5  for (a) REM and (b) DETREM simulations.}
\end{figure}

Figure \ref{ym} shows the time series of $y_m$ change in one of the temperature pairs $(m=1)$ as a function of MC steps from the DETREM simulation. We see a random walk, which results in random walks in temperature space for each replica.
Figure \ref{rw}(a) and Figure \ref{rw}(b) show the time series of temperature change in one of the replicas (Replica 1) as a function of MC steps from the conventional REM and the DETREM simulation, respectively.
They show similar behaviors with respect to random walks in temperature space. 
We see that all replicas take the minimum temperature many times during both simulations.
Other replicas perform random walks similarly.
Figures \ref{ind}(a) and \ref{ind}(b) show the time series of replica index at the minimum temperature of 1.5 during the REM and DETREM simulations.
This shows that all replicas experienced the minimum temperature many times during the simulation.
Table \ref{tc} lists the maximum number of  tunneling events per replica, which is the number of times where the simulation visits from the lowest temperature through the highest temperature and back to the lowest temperature. 
These data show that the two methods have nearly the same number of  tunneling counts during the simulations. 
All these results imply that REM and DETREM are equally efficient in sampling.

\begin{figure}[htb]
\centering
\includegraphics[width=1.0\textwidth]{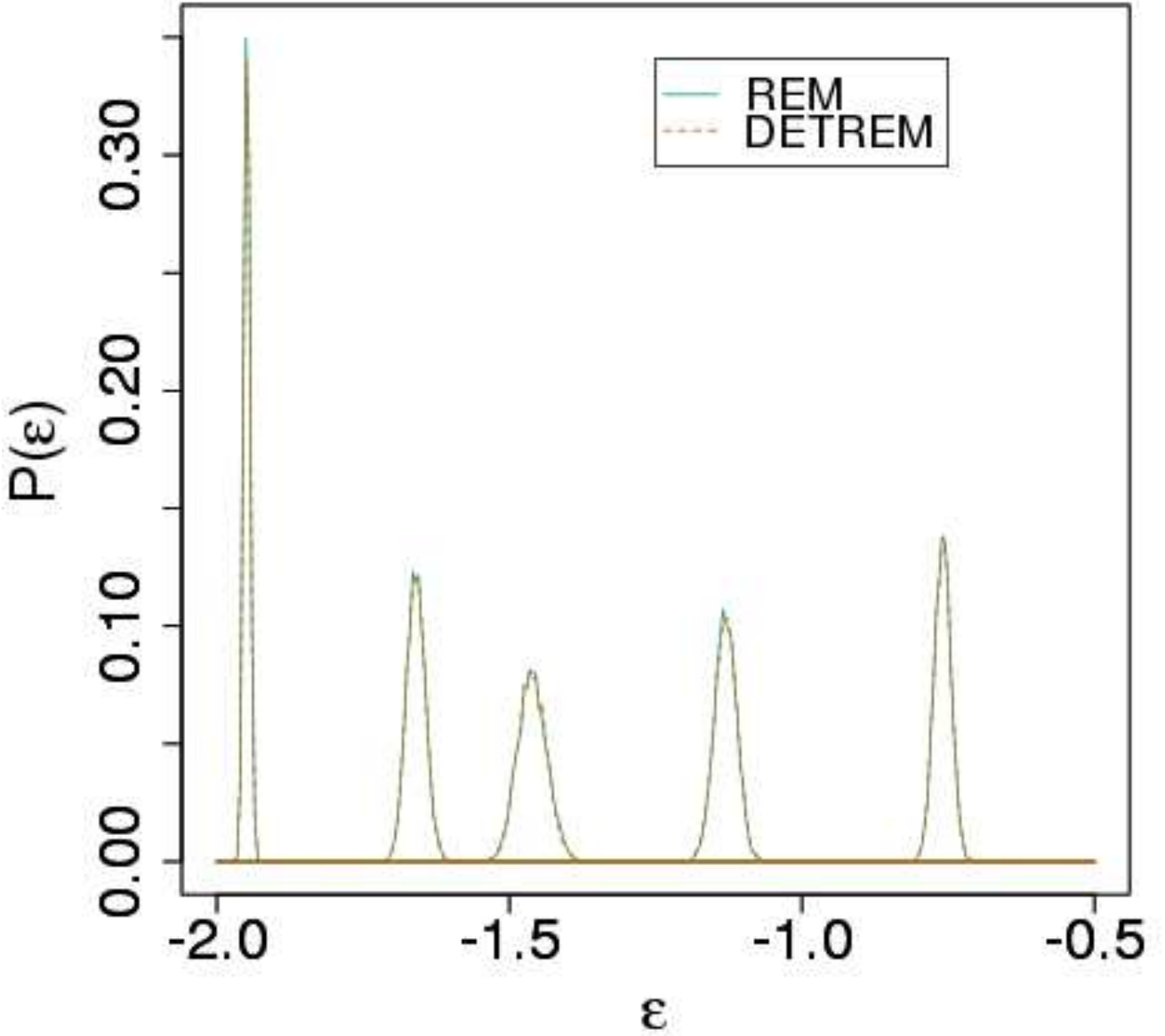}
\caption{\label{dist}Probability distributions of energy density at four temperatures (from left to right, 1.50, 2.10, 2.25, 2.47, and 3.15) obtained from (a) REM and (b) DETREM simulations including the mixed-walk simulation.}
\end{figure}

We next examine physical quantities obtained from the DETREM simulation and compare them to those from the REM simulation.
Figure \ref{dist} shows the canonical energy distributions at 40 temperatures as functions of energy obtained from the REM and DETREM simulations.
We see that the distributions have enough overlaps in pairs of the neighboring distributions. 
 This ensures that the number of replicas is sufficient.
The agreement between two methods implies that DETREM method produced the Boltzmann distributions at each temperature  simulation just like REM did.

\begin{figure}[htb]
\centering
\includegraphics[width=1.0\textwidth]{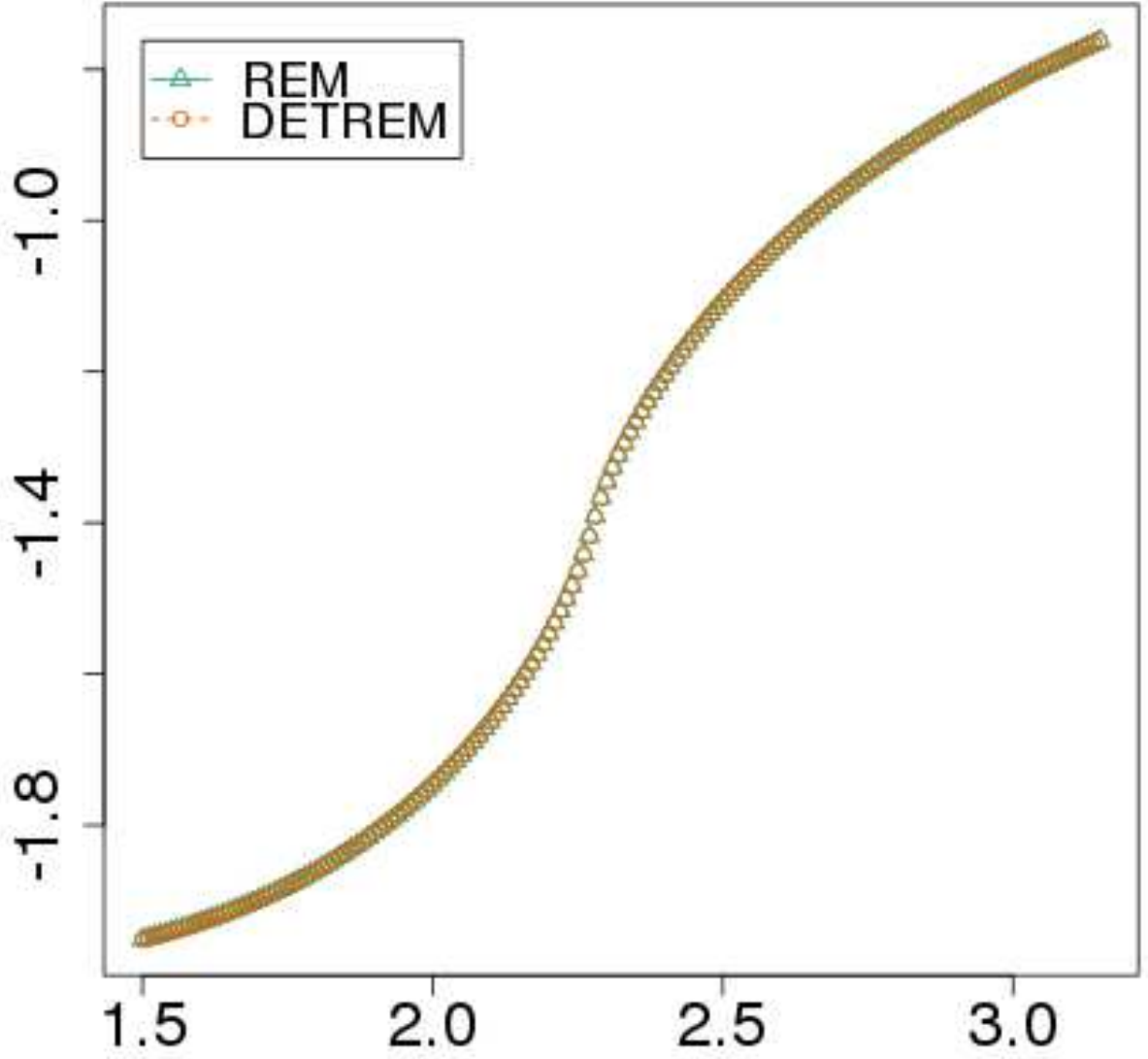}
\caption{\label{energy}Average energy density as a function of temperature obtained by WHAM from the REM and DETREM simulations. The error bars are smaller than the symbols.}
\end{figure}

\begin{figure}[htb]
\centering
\includegraphics[width=1.0\textwidth]{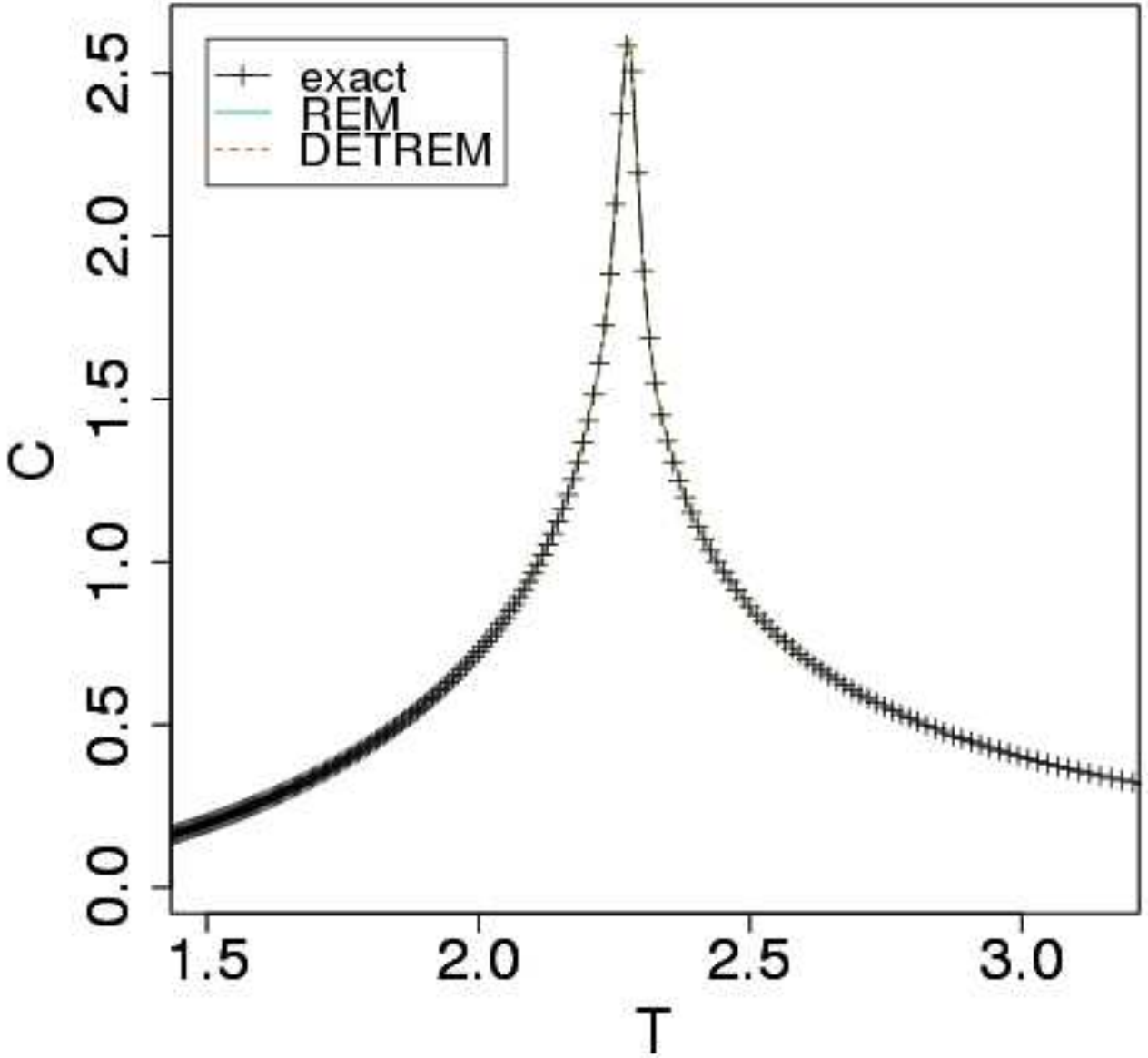}
\caption{\label{capa}Specific heat as a function of temperature obtained by WHAM from the REM and DETREM simulations. The error bars are smaller than the symbols. The exact results for $L$ =128 (black curves) were obtained by Berg's program \cite{berg2004markov} based on Ref. \cite{ferdinand1969bounded}.}
\end{figure}

We next confirm the second-order phase transitions at the critical temperature of $T_{\rm c}\sim2.269$ in both methods.
 Figure \ref{energy}(a) and Figure \ref{energy}(b) show the total energy density $\epsilon$ as a function of $T$ during the REM simulation and the DETREM simulation, respectively, where $\epsilon$ is defined by
\begin{equation}
\displaystyle \epsilon =\frac{E} { N}.
\end{equation}
Figure \ref{capa}(a) and Figure \ref{capa}(b) show the specific heat $C$ as a function of $T$ from the REM simulation and the DETREM simulation, respectively, where $C$ is defined by 
\begin{equation}
\displaystyle C = \frac{1} {k T^2 N} (<E^2> - <E>^2).
\end{equation}

\begin{figure}[htb]
\centering
\includegraphics[width=1.0\textwidth]{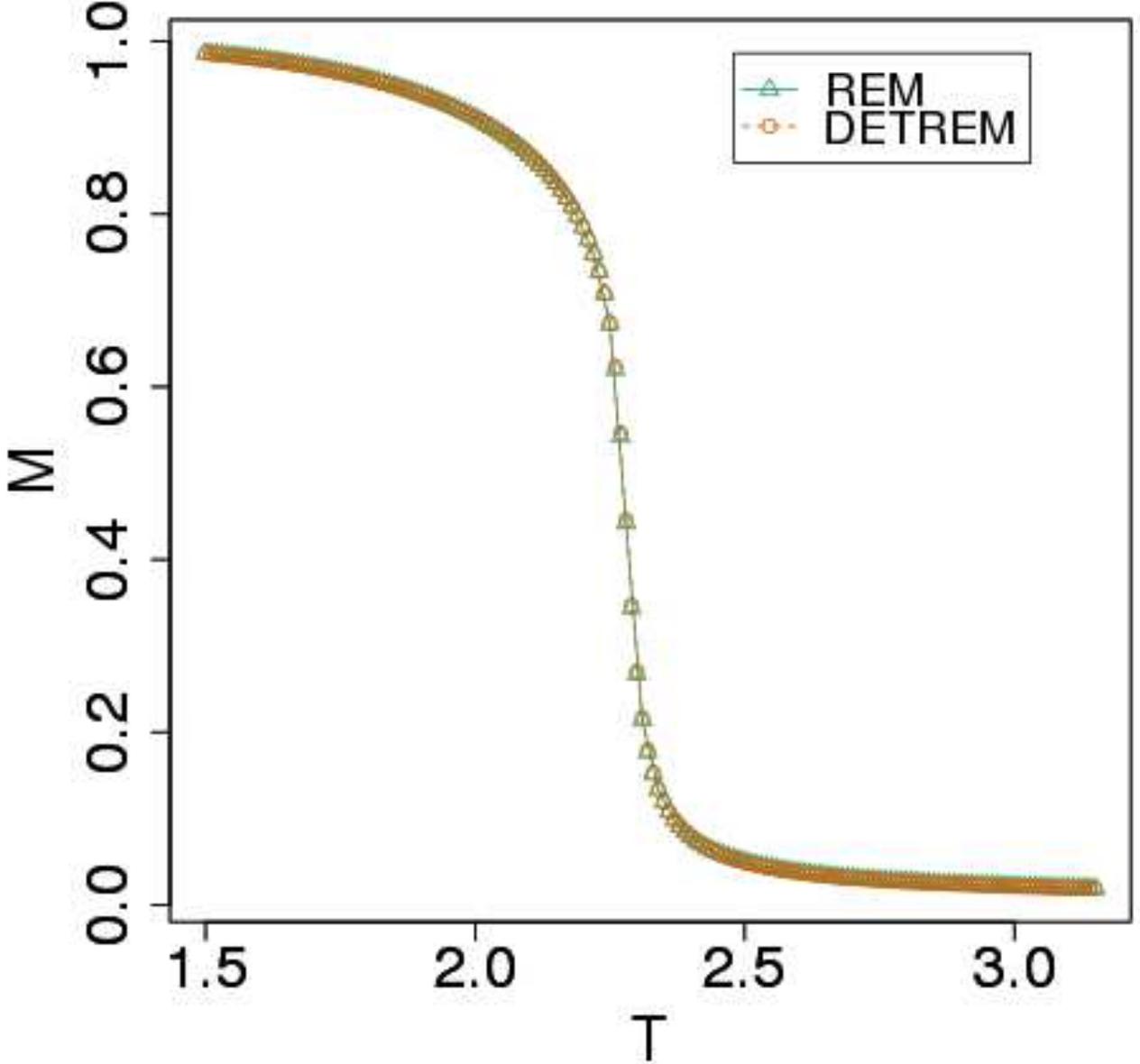}
\caption{\label{mag}Magnetization as a function of temperature obtained by WHAM from the  REM and  DETREM simulations. The error bars are smaller than the symbols.}
\end{figure}

\begin{figure}[htb]
\centering
\includegraphics[width=1.0\textwidth]{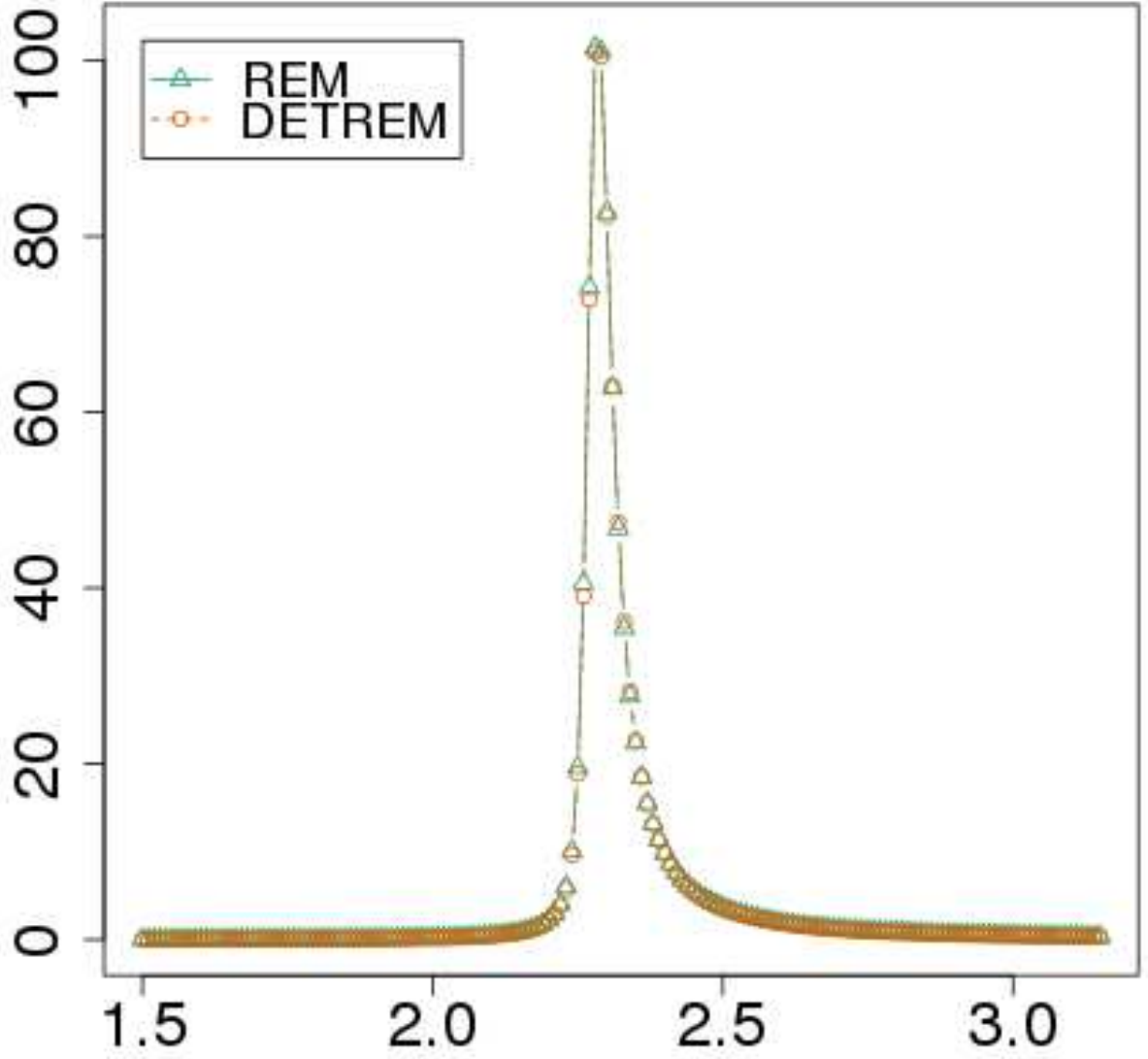}
\caption{\label{xi}Susceptibility as a function of temperature obtained by WHAM from the REM and DETREM simulations.  The error bars are smaller than the symbols.}
\end{figure}

Figure \ref{mag}(a) and Figure \ref{mag}(b) show magnetization $M$ as a function of $T$ from the REM simulation and the DETREM simulation, respectively, where $M$ is defined by
\begin{equation}
\displaystyle M =\frac{\displaystyle  \Biggl| \sum_{i=1}^N s_i \Biggr|}{ N}.
\end{equation}
Figure \ref{xi}(a) and Figure \ref{xi}(b) show susceptibility $\chi$ as a function of temperature during the REM simulation and the DETREM simulation, respectively, where $\chi$ is defined by
\begin{equation}
\displaystyle \chi =  \frac{N} {k T } (<M^2> - <M>^2). 
\end{equation}

\begin{figure}[htb]
\centering
\includegraphics[width=1.0\textwidth]{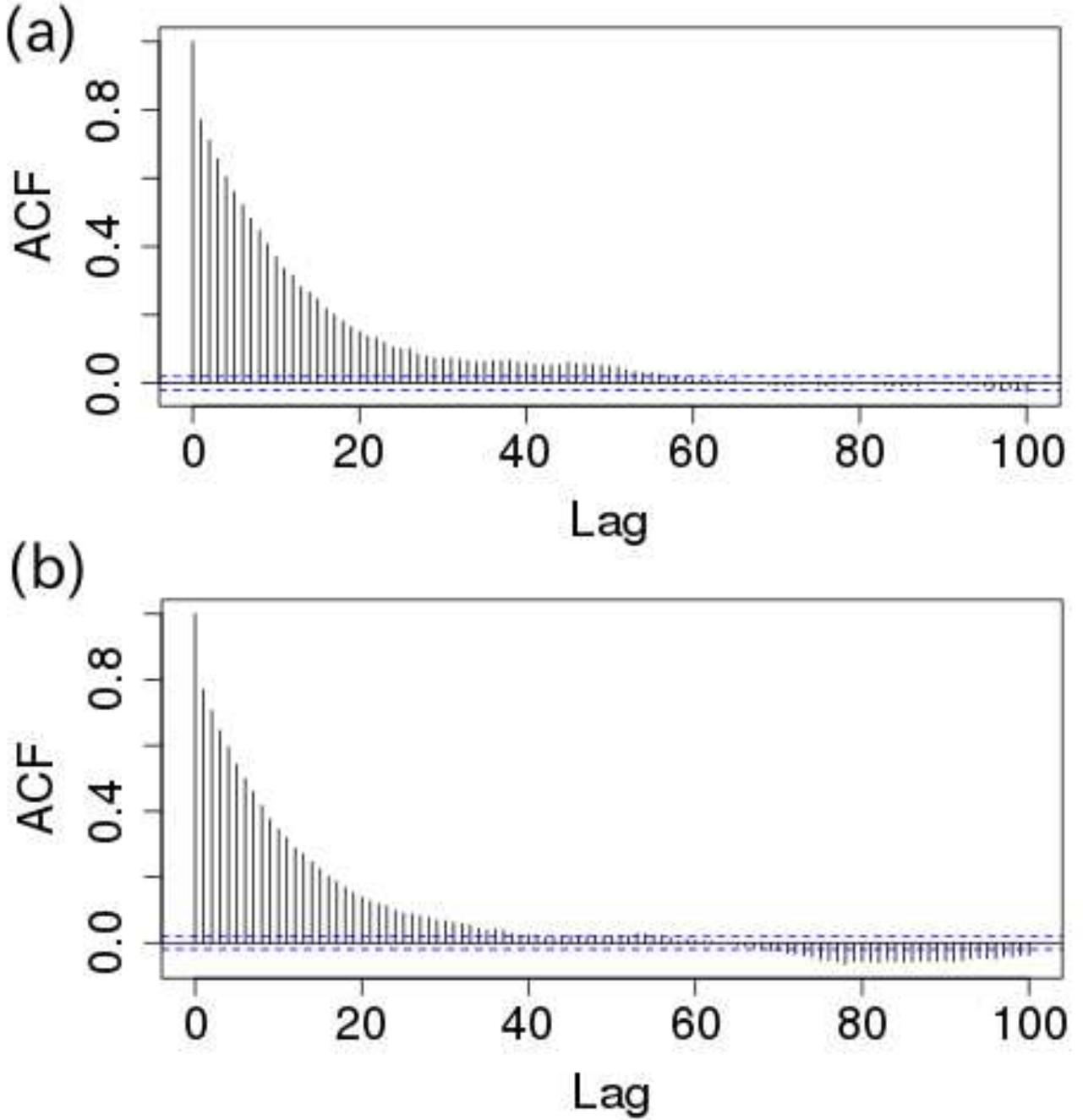}
\caption{\label{acf_ene}Autocorrelation function of total energy density as a function of MC step lag in one of  the replicas (Replica 1) for (a) REM and (b) DETREM simulations. The blue dashed lines show a 95=\% confidence interval.}
\end{figure}

Figure \ref{acf_ene}(a) and Figure \ref{acf_ene}(b) show the autocorrelation function of total energy density  as a function of MC step lags from the REM simulation and the DETREM simulation, respectively. Here, the autocorrelation function of a physical quantity $A$ is defined by
    \begin{eqnarray}
      C(t) &=& \frac{<A(0) A(t) > - <A>^2}{<A^2> - <A>^2}  \\  
    &=&  \frac{1}{n} \sum_{s=max(1,-t)}^{min(n-t,n)} <A_{s+t} -
        <A> > <A_s - <A>>,  
    \end{eqnarray}
    where the average $<\cdots > $ is over the samples, $t$ is the MC step lag, and the
    second line is written for discrete steps ($A_t $ for $t=1, 2, \cdots, n$) from the $n-t$
    observed pairs $(A_{1+t},A_1),\cdots , (A_n, A_{n-t})$, 
The decay coefficient  $\tau  \ (\rm in\  C(t) \propto e^{\tau t})$ is determined from a linear fit of log $C(t))$ to lag
 in the region where the long-time single exponential decay is observed.
This shows that the autocorrelation function of total energy density behaves similarly between REM and DETREM, and the decay coefficient $\tau$ of REM and DETREM in the Figures is  
  $-0.066\pm0.002$ and $-0.080\pm  0.003$, respectively. 

\begin{figure}[htb]
\centering
\includegraphics[width=1.0\textwidth]{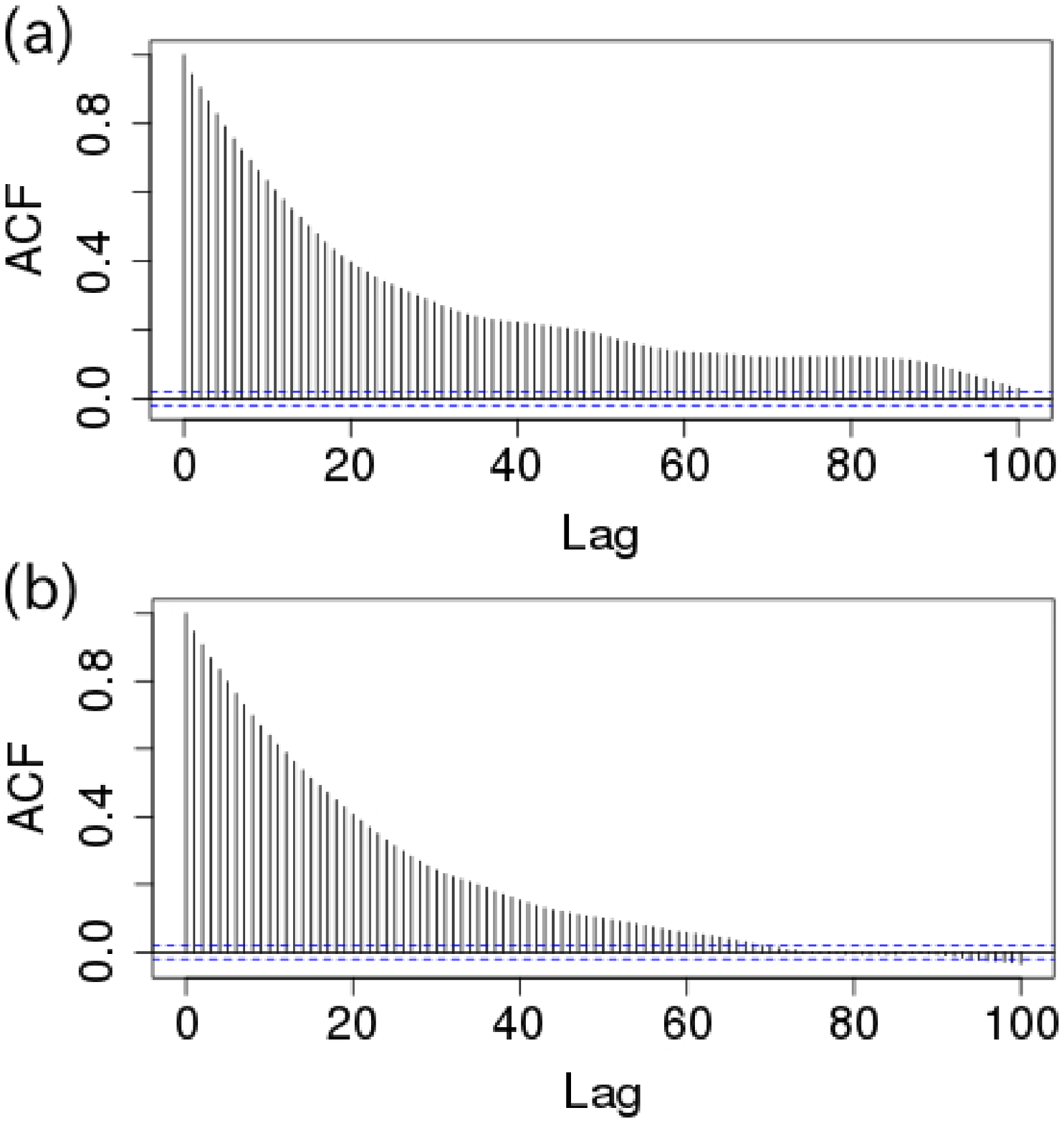}
\caption{\label{acf_mag}Autocorrelation function of magnetization  as a function of MC step lag in one of  the replicas (Replica 1) for (a) REM and (b) DETREM simulations.  The blue dashed lines show a 95=\% confidence interval.}
\end{figure}

Figure \ref{acf_mag}(a) and Figure \ref{acf_mag}(b) show the autocorrelation function of magnetization $M$  as a function of MC step lags   from the REM simulation and the DETREM simulation, respectively.
This shows that the autocorrelation of DETREM decreases faster than that of REM and that the decay coefficient $\tau$ of REM and DETREM is
$-0.025\pm 0.001$ and $-0.057\pm   0.002$, respectively. 
Thus, these results suggest that the autocorrelation for total energy density and magnetization in DETREM both decreases  faster than that in REM. 

All these physical quantities (Figs. \ref{dist} to \ref{acf_mag} ) confirm that the DETREM simulation not only reproduced the results of the REM simulation in the phase transitions near the critical temperature $T_{\rm c}$ but also with faster convergence.

\section*{Conclusions}
\label{sec-4}
In this work, we proposed a deterministic replica-exchange method, which enables replicas to exchange their temperatures keeping their thermal equilibrium without using pseudo random numbers.
We reproduced the results of REM by DETREM. 
This fact may be useful for parallel computing because the correlation in pseudo random number sequences in REM may cause defects in the results of MC simulations\cite{li2009decentralized}.  For example, some MC simulations which include association in sites may include hidden errors in their results\cite{ferrenberg1992monte}.  Moreover, for molecular dynamics simulations, the use of bad seeds in stochastic thermostat caused the partial synchronization of trajectories\cite{sindhikara2009bad}. 
Thus, similar synchronization of trajectories may occur by replica exchange with  pseudo random numbers.
Although to the best of our knowledge no work directly observing this phenomenon exists, we may encounter with this problem in the future.
 Moreover, DETREM can be considered to be a special application of the 
 deep Boltzmann 
 machine\cite{salakhutdinov2009deep,salakhutdinov2012efficient} in 
 machine learning.
 Thus, various effective method in machine learning can be introduced 
 in order to further enhance the performance of DETREM.
In addition, this new method can also be applied the transformation of conditional probability to other replica-exchange method variants.
In another future work, we will introduce the multidimensional DETREM for generalized potential function including Hamiltonian replica-exchange method\cite{sugita2000multidimensional,fukunishi2002hamiltonian}.

\section*{Acknowledgments}
\label{sec-5}
We are grateful to Drs. Yoshiharu Mori and Tetsuro Nagai for informing us the existence of Ref. \cite{chodera2011replica}.
Some of the computations were performed on the supercomputers at the
Institute for Molecular Science, at the Supercomputer Center, Institute
for Solid State Physics, University of Tokyo, and Center for
Computational Sciences, University of Tsukuba.
This work was supported, in part, Grants-in-Aid for Scientific Research (A) (No. 25247071),
for Scientific Research on Innovative Areas (\lq\lq Dynamical Ordering \& Integrated
Functions\rq\rq ), Program for Leading Graduate Schools \lq\lq Integrative Graduate Education and Research in Green Natural Sciences\rq\rq, and for the Computational Materials Science Initiative, for High Performance Computing Infrastructure, and CREST
"Molecular Technology for Chemical Control of Epigenetics
towards Drug Discovery" from the Ministry of Education,
Culture, Sports, Science and Technology (MEXT), Japan and
Japan Science and Technology Agency (JST).

\begin{table}[!tbp]
\caption{The number of maximum tunneling count (TC) per replica of whole replicas during simulations\label{tc}} 
\begin{center}
\begin{tabular}{l|r|r}
\hline\hline
\multicolumn{1}{l|}{TC}&\multicolumn{1}{|c|}{REM}&\multicolumn{1}{|c}{DETREM}\tabularnewline
\hline
Max &$191 $&$ 196 $\tabularnewline 
\hline
Mean $\pm$ SD &$173.3\pm 9.5 $ &$ 177.95\pm  8.8$\tabularnewline 
\hline
\hline
\end{tabular}
\end{center}
SD means standard deviation with respect to replicas.
\end{table}

\appendix

\section{General Formalism of Related Methods}
We present the equations for the general formalism
of related methods in the following.

\subsection{GSREM}
For Gibbs sampling replica-exchange method, general equations are given in Ref. \cite{chodera2011replica}.
For REM with heat bath method, the conditional probability assigned for new states for any replica-exchange is given by
\begin{eqnarray}
\label{gcond}
\displaystyle \omega(S \mid x^{[1]}, x^{[2]}, \cdots, x^{[M]})  
&= &\cfrac{ 
W(x_{m(1)}^{[1]}, x_{m(2)}^{[2]}, \cdots, x_{m^(M)}^{[M]}) 
}{
\displaystyle \sum_{S^\prime \in S_M } W(x_{m^\prime(1)}^{[1]}, x_{m^\prime(2)}^{[2]}, \cdots, x_{m^\prime(M)}^{[M]})
}, 
\end{eqnarray}
where $S\equiv \left\{{m(1)},{m(2)}, \cdots,{m(M)}\right\}$ is a permutation of temperature indices and $S_M$ is all possible permutations.

When we consider exchange of only one pair of replicas to reduce the set $S_M$, Eq. (\ref{gcond}) turns into 
\begin{eqnarray}
\label{allpair}
\displaystyle \omega(x_{m(i)}^{[i]}, x_{m^(j)}^{[j]} \mid x^{[k \ne m(i),m(j)]}_{m(k)})  
&= &\cfrac{ 
W(x_{m(i)}^{[i]}, x_{m(j)}^{[j]}) 
}{
\displaystyle \sum_{i^\prime =1 }^{M-1}\sum_{j^\prime >i^\prime }^{M} W(x_{m(i^\prime)}^{[i^\prime]}, x_{m(j^\prime)}^{[j^\prime]})
}. 
\end{eqnarray}

\subsection{DETREM}
We propose the differential equations for DETREM corresponding to the general equations of GSREM.
At first, we introduce an internal state $y$ to assign a permutation of temperature state based on  Eq. (\ref{gcond}).
It satisfies the following differential  equation: 
\begin{eqnarray}
\label{ypermu}
\displaystyle \frac{d y }{dt}
&= &\cfrac{ 
W(x_{m(1)}^{[1]}, x_{m(2)}^{[2]}, \cdots, x_{m^(M)}^{[M]}) 
}{
\displaystyle \sum_{S^\prime \in S_M } W(x_{m^\prime(1)}^{[1]}, x_{m^\prime(2)}^{[2]}, \cdots, x_{m^\prime(M)}^{[M]})
}, 
\end{eqnarray}
where $t$ is a virtual time, $y \in [1,N_{S_M}+1]$, and $N_{S_M}$ is the total number of elements in the permutation set $S_{M}$. 
This DETREM is performed just like GSREM, where in Step 2 the evaluation of the conditional probability in Eq. (\ref{gcond}) is replaced by solving the differential equation in Eq. (\ref{ypermu}). DETREM gives the same probability as in Eq. (\ref{gcond}). 
 When $y$ reaches $m=\lfloor y\rfloor$, the permutation of temperature corresponding to the integer in temperature permutations is chosen. 
Here, the floor function $\lfloor y \rfloor$ is the integral part of $y$ defined by $\lfloor y \rfloor = m\leftrightarrow m \leq y < m+1$.
In general, $N_{S_M}$ is so large that using Eq. (\ref{ypermu}) is not practical.  

We thus introduce an internal state $y_{m,n}$ for exchange of a pair of temperatures  $T_m$ and $T_n$ that correspond to replicas $i$ and $j$, respectively, with the time integration based on Eq. (\ref{allpair}), which gives 
\begin{eqnarray}
\displaystyle \frac{d y_{m,n} }{dt}
&= &\cfrac{ 
W(x_{m}^{[i]}, x_{n}^{[j]}) 
}{
\displaystyle  W(x_{m}^{[i]}, x_{n}^{[j]})  +W(x_{n}^{[i]}, x_{m}^{[j]}) 
}\\
&=& \frac{1}{1+ \rm exp(\Delta_{m,n} )},
\end{eqnarray}
where $y_{m,n} \in [0,1]$ and $\Delta_{m,n}$ is given by Eq. (\ref{ddelta}).

\subsection{DETST}
We remark that simulated tempering (ST)\cite{lyubartsev1992new,marinari1992simulated} corresponding to GSREM and DETREM  can also be formulated. 
The weight factor for ST is given by
\begin{eqnarray}
W_m^{\rm ST}(x)&=\displaystyle  \exp{[-\beta_{m} E(x) +f_m]}, 
\end{eqnarray}
where $f_m$ are the dimensionless Helmholtz free energy at temperature $T_m \ (m=1, \cdots,M)$.
In Step 1, we perform a canonical MC or MD simulation at temperature $T_m$ for short steps.
In Step 2, when we consider temperature change to neighboring values\cite{shida2006gibbsst}, the conditional probability from temperature $T_m$ into $T_{m+1}$ or $T_{m-1}$ is given by 
\begin{eqnarray}
\displaystyle \omega(T_{m \pm 1}  \mid  T_{m} ,x)  
&= &\cfrac{ \displaystyle W^{\rm ST}_{m \pm 1}  (x)  }{ \displaystyle
W^{\rm ST}_{m \pm 1} (x) + W^{\rm ST}_{m }(x)
}
= \cfrac{ 1}{\displaystyle 1  + \cfrac{W^{\rm ST}_{m }(x)} {W^{\rm ST}_{m \pm 1} (x)}} \\
& = & \cfrac{1}{1+ \Delta^{\rm ST}_\pm}\ , 
\end{eqnarray}
where $\Delta^{\rm ST}_\pm$ are defined by 
\begin{eqnarray}
\Delta^{\rm ST}_\pm = ( \beta_{m \pm 1} - \beta_m ) E(x) - (f_{m \pm 1} - f_m).
\end{eqnarray}
This is the transition probability for the Gibbs sampling simulated tempering (GSST) (this formulation was first given in Ref. \cite{chodera2011replica}).
Hence, the internal states and differential equations are given by 
\begin{eqnarray}
\frac{dy_m^+}{dt} =    \cfrac{1}{1+ \Delta^{\rm ST}_+}\ ,      \\
\frac{dy_m^- }{dt}=\cfrac{1}{1+ \Delta^{\rm ST}_-}\ ,   
\end{eqnarray}
where $y_m=y_{m,m+1}\ \ \ (m= 1, \cdots, M-1)$. 
 When the system stays at temperature $T_m$, only $y_m^\pm$ are updated, and other $y_{n\ne m}^\pm$ are not updated.

New temperature is updated to a new value $T_n$ with the following conditional probability:
\begin{eqnarray}
\omega(T_n \mid x) &= & \cfrac{ 
W_n^{\rm ST}(x)
}{ \displaystyle 
\sum_{m =1 }^{M} W_m^{\rm ST}(x)
} =\cfrac{
\exp{[- \beta_n E(x)+ f_n]}
}{\displaystyle 
\sum_{m =1 }^{M}  \exp{[- \beta_{m} E(x)+ f_{m}]}
} \\
& \propto & \exp{[ - ( \beta_{n} -  \beta_{0}) E(x) + (f_{n}-f_{0}) ]},
\end{eqnarray}
where we have introduced an arbitrary reference temperature $T_0$ to give the normalization.

From above, we can derive a differential equation for deterministic simulated tempering (DETST).
We introduce an internal state $y$
\begin{eqnarray}
\cfrac{ dy}{dt}  = \exp{[ - ( \beta_{n} -  \beta_{0}) E(x) + (f_{n}-f_{0}) ]},
\end{eqnarray}
where $y \in \left\{1, \cdots, M+1 \right\}$ and $m=\lfloor y\rfloor$. 

In another implementation, the conditional probability  from the current temperature $T_m$ to $T_n$ is given by
\begin{eqnarray}
\omega(T_n \mid x,T_m)
= \cfrac{ \displaystyle W^{\rm ST}_{n}  (x)  }{ \displaystyle
W^{\rm ST}_{n} (x) + W^{\rm ST}_{m }(x)
}\ .
\end{eqnarray}

Thus, we introduce an internal state $y_{m,n}$ integrated by 
\begin{eqnarray}
\cfrac{ dy_{m,n}}{dt}  
&= &\cfrac{ \displaystyle W^{\rm ST}_{n}  (x)  }{ \displaystyle
W^{\rm ST}_{n} (x) + W^{\rm ST}_{m }(x)
}
= \cfrac{ 1}{\displaystyle 1  + \cfrac{W^{\rm ST}_{m }(x)} {W^{\rm ST}_{n} (x)}} \\
& = & \cfrac{1}{1+ \Delta^{\rm ST}_{m,n}}\ , 
\end{eqnarray}
where $y_{m,n} \in \left\{0,1\right\}$, $m,n \in \left\{1, 2, \cdots, M\right\}$, and 
\begin{eqnarray}
\Delta^{\rm ST}_{m,n} =( \beta_{n} - \beta_m ) E(x) - (f_{n} - f_m).
\end{eqnarray}

\bibliographystyle{apsrev4-1}
\bibliography{ref}

\end{document}